\documentclass[aps]{revtex4}
\usepackage{graphicx}% Include figure files
\usepackage{dcolumn}% Align table columns on decimal point
\usepackage{bm}% bold math

\begin{document}
\hfill{LAPTH-1025/04, IFIC/04-08}

\title{Current cosmological bounds on neutrino masses and
  relativistic relics}

\author{Patrick Crotty} \altaffiliation{Present address: Department of
  Neurosurgery, University of Virginia Health System, PO Box 800420,
  Charlottesville, VA 22908, USA}
\author{Julien Lesgourgues}
\affiliation{Laboratoire de Physique Th\'eorique LAPTH, B.P. 110,
 F-74941 Annecy-le-Vieux Cedex, France}
\author{Sergio Pastor}
\affiliation{Instituto de F\'{\i}sica Corpuscular (CSIC-Universitat de
Val\`encia), Ed.\ Institutos de Investigaci\'on, Apdo.\ 22085,
E-46071 Valencia, Spain}

\date{February 4, 2004}

\begin{abstract}
  We combine the most recent observations of large-scale structure
  (2dF and SDSS galaxy surveys) and cosmic microwave anisotropies
  (WMAP and ACBAR) to put constraints on flat cosmological models where
  the number of massive neutrinos and of massless relativistic relics
  are both left arbitrary.  We discuss the impact of each dataset and
  of various priors on our bounds. For the standard case of three
  thermalized neutrinos, we find $\sum m_{\nu} < 1.0 \, ({\rm resp.}
  \, 0.6)$ eV (at 2$\sigma$), using only CMB and LSS data (resp.
  including priors from supernovae data and the HST Key Project), a
  bound that is quite insensitive to the splitting of the total mass
  between the three species. When the total number of neutrinos or
  relativistic relics $N_{\rm eff}$ is left free, the upper bound on
  $\sum m_{\nu}$ (at 2$\sigma$, including all priors) ranges from
  $1.0$ to $1.5$ eV depending on the mass splitting. We provide an
  explanation of the parameter degeneracy that allows
  larger values of the masses when $N_{\rm eff}$ increases. Finally,
  we show that the limit on the total neutrino mass is not
  significantly modified in the presence of primordial gravitational
  waves, because current data provide a clear distinction between the
  corresponding effects.
\end{abstract}

\maketitle

\section{Introduction}

Neutrino properties are among the most difficult to be probed
experimentally, due to the weakness of their interactions.  {}Data
from particle accelerators tell us that there are only three flavor
neutrinos, while neutrino oscillation experiments show evidence for
non-zero neutrino masses (for a recent review, see e.g.\ 
\cite{Maltoni:2004ex}). Recent results strongly constrain the mass
differences of the individual neutrino masses (actually masses
squared, $\Delta m^2$) and mixing angles, but no definite conclusion
can be drawn neither on the absolute scale of neutrino masses, nor on
the existence of weakly coupled sterile neutrinos.  Fortunately,
cosmology is quite sensitive to the neutrino sector (see
\cite{Dolgov:2002wy} for a review), and can shed light on these
questions, as well as other interesting issues regarding the Early
Universe, such as the process of neutrino decoupling from the
primordial plasma.

Currently, the most popular cosmological model is the flat adiabatic
$\Lambda$CDM scenario, in which the present density of the Universe is
shared between baryons, Cold Dark Matter (CDM) and a cosmological
constant $\Lambda$. This model makes rather simplistic assumptions
concerning the neutrino sector, consisting only of three
ultra-relativistic neutrinos. It turns out that with a more refined
description of the neutrino sector, one finds that only small
corrections to the standard picture are allowed after comparing with
current data on Cosmic Microwave Background (CMB) anisotropies and
Large Scale Structure (LSS).  However, these small corrections carry
enough interesting physical implications to justify an active research
effort, in particular after the first releases of the WMAP and SDSS
data. The results of this effort are not only new cosmological bounds
on neutrino properties but also a better understanding of how the
errors depend on (i) the experimental CMB and LSS data, (ii) external
priors on the cosmological parameters, (iii) intrinsic parameter
degeneracies in the theory of cosmological perturbations, (iv)
assumptions concerning the underlying cosmological model and parameter
space.

We here perform a new analysis using the most recent LSS (2dF, SDSS)
and CMB (WMAP, ACBAR) data and an extended cosmological model with an
arbitrary number of massive neutrinos and additional relativistic
particles, parametrized via an effective number of neutrinos ($N_{\rm
  eff}$). We extend the recent work of \cite{Barger:2003vs} and those
that appeared after the release of WMAP data
\cite{Spergel:2003cb}-\cite{Hannestad:2003ye}.
%\cite{Spergel:2003cb,Crotty:2003th,
%  Pierpaoli:2003kw,Hannestad:2003xv,Elgaroy:2003yh,Barger:2003zg,
%  Allen:2003pt,Cuoco:2003cu,Tegmark:2003ud,Hannestad:2003ye}.  
In particular, our underlying model is identical to that of ref.\ 
\cite{Hannestad:2003ye}, but our analysis differs since we include an
extended set of data (such as the SDSS results and a more updated
version of the 2dF ones), a new prior on the matter density from SN-Ia
\cite{Tonry:2003zg} and non-linear corrections to the LSS power
spectrum on scales $0.1 \, h~{\rm Mpc}^{-1} < k < 0.2 \, h~{\rm
  Mpc}^{-1}$.  Furthermore, we increase the number of free parameters
to ten, in order to analyze the bounds in the presence of primordial
tensor perturbations.

The rest of the paper is organized as follows. After a short summary
of the effects of neutrino masses and additional relativistic
particles in Sec.\ \ref{effects}, we describe our analysis method and
dataset in Sec.\ \ref{method_data}. We discuss our results and compare
with previous works in Sec.\ \ref{results}. Finally, we conclude in
Sec.\ \ref{conc}.

\section{Effects of additional relativistic particles and massive neutrinos}
\label{effects}

Non-standard neutrinos and other weakly-interacting light particles
leave their imprint on the evolution of the Universe, both at the
level of background quantities and spatial perturbations. Here we
describe the main effects of additional relativistic particles,
massive neutrinos and their simultaneous presence.

\subsection{Enhanced relativistic energy density ($N_{\rm eff}$)}

The density of radiation in the Universe is usually assumed to be
given by that of photons and of three thermally decoupled neutrinos.
These contributions are of the same order and fix the evolution of the
Universe in the radiation-dominated epoch (RD).  Thus, if the three
neutrinos did not decouple thermally, or in the presence of sterile
neutrinos, the total density of the Universe during RD (as a function
of the photon temperature $T_{\gamma}$) would be significantly
affected, producing a change in the time of equality between radiation
and matter, and in the sound horizon at the time of decoupling.  These
changes are known to shift the angular scale of the acoustic peaks in
the CMB anisotropy spectrum as well as their amplitude (mainly,
through the early integrated Sachs-Wolfe effect).  They also have an
impact on the matter power spectrum $P(k)$, because a shorter
matter-dominated stage implies less growth for perturbations inside
the Hubble radius.  As a consequence, the wave-length corresponding to
the maximum in $P(k)$ will be shifted proportionally to the Hubble
scale at the time of equality. Thus, the effect of $N_{\rm eff}$ is
mainly to change the background evolution. However, ultra-relativistic
particles also have a smaller effect directly at the level of
perturbations, explained in detail in Ref.~\cite{Bashinsky:2003tk}.

All these effects can be parametrized by a single quantity: the effective
number of relativistic degrees of freedom during RD,
defined by the relation
\begin{equation}
\rho_r = \left[ 1 + \frac{7}{8} \left( \frac{4}{11} \right)^{4/3} 
N_{\rm eff} \right] \frac{\pi^2}{15} T_{\gamma}^4 \, .
\label{def_Neff}
\end{equation}
Here, $\rho_r$ stands for the total energy density of radiation and
$\rho_\gamma=(\pi^2/15) \, T_{\gamma}^4$ is the contribution of photons.
The parameter $N_{\rm eff}$ is defined in such way that if neutrinos
decoupled following the instantaneous decoupling approximation,
$N_{\rm eff}$ just gives the number of flavor families.  However,
$N_{\rm eff}$ could differ from three in the presence of extra relics
(sterile neutrinos, light gravitons, gravitinos, majorons, effects
from extra dimensions, etc.) or in the case of non-thermal decoupling.
Actually, in the standard case a careful study of non-instantaneous
neutrino decoupling shows that $N_{\rm eff}=3.04$ for three flavor
families \cite{Dolgov:1997mb, Mangano:2001iu}. Note that $N_{\rm eff}$
is constant only when the neutrinos or the other relics are
ultra-relativistic.

The value of $N_{\rm eff}$ is constrained by Big Bang Nucleosynthesis
(BBN) from the comparison with the measured primordial abundances of
light elements. During the BBN epoch the nuclear reactions freeze out
at a scale factor that depends on the expansion rate, which in turn is
fixed by the total energy density of radiation. A BBN analysis shows
that $N_{\rm eff}= 2.5^{+1.1}_{-0.9}~ (2\sigma)$ \cite{Cuoco:2003cu}
(see also \cite{Cyburt:2003fe}), which is perfectly compatible with
the number of flavor neutrinos.

However, it is interesting to measure $N_{\rm eff}$ independently of
BBN (e.g. using CMB and LSS data) for at least two reasons. First of
all, because the number $N_{\rm eff}$ could change between the two
epochs \cite{White:1994as,Kaplinghat:2000jj}.  A second reason is
because the standard BBN model might be a good first-order
description, but with possible corrections due to spatial
inhomogeneities, leptonic asymmetries, etc., that could be evaluated
with an independent measurement of $N_{\rm eff}$.

\subsection{Massive neutrinos}

Neutrinos that possess masses larger or of the same order than the
relevant photon temperature have different effects than a constant
$N_{\rm eff}$.  For instance, neutrinos heavier than roughly $10^{-3}$
eV are not relativistic today. Neutrino masses have implications for
the evolution of cosmological fluctuations, both at the level of
background quantities and directly on the perturbations.

It is well-known that massive neutrinos could account for a
significant fraction of the total energy density of the Universe
today, unlike relativistic thermal relics ($\Omega_r \sim 5.6 \times
10^{-6}$ per neutrino family). For fully non-relativistic flavor
neutrinos, the contribution to the present energy density is directly
proportional to the number density. For vanishing neutrino chemical
potentials, the total neutrino contribution to the critical density is
given by
\begin{equation}
\Omega_\nu =  \frac{\sum m_\nu}{93.2~{\rm eV}} \, h^{-2}~,
\label{omeganuh2}
\end{equation}
where $h$ is the Hubble constant in units of $100$ km s$^{-1}$
Mpc$^{-1}$ and $\sum m_\nu$ runs over all neutrino mass states.  For
fixed neutrino masses, $\Omega_\nu$ would be enhanced if neutrinos
decoupled with a significant dimensionless chemical potential $\xi_\nu
\equiv \mu_\nu/T$ (or equivalently, for large relic neutrino
asymmetries), simply because their number density would increase. In
principle there exist some combinations of pairs
$(\xi(\nu_e),\xi(\nu_{\mu,\tau}))$ that pass the BBN test and are
not yet ruled out by the CMB+LSS limits on $N_{\rm eff}$
\cite{Lesgourgues:1999wu}. However, it was recently shown that the
stringent BBN bounds on $\xi_e$ apply to all flavors, since flavor
neutrino oscillations lead to flavor equilibrium before BBN
\cite{Dolgov:2002ab,Wong:2002fa,Abazajian:2002qx}. The contribution of
a potential relic neutrino asymmetry is limited to such low values
that it can be safely ignored.

When the density of the other fluids (photons, CDM, baryons, dark
energy) is kept fixed, the sum over the neutrino masses $\sum m_{\nu}$
has a direct repercussion on the geometry of the Universe.  If instead
the spatial curvature is kept fixed, the total mass affects the
relative contribution $\Omega_X$ of the other fluids. In any case,
this background effect has an impact on the observable CMB and LSS
power spectra. For masses of the order of $1$ eV, this signature is
rather small, but can be marginally detectable.

In general, neutrinos tend to stream freely across gravitational
potential wells, and to erase density perturbations. Free-streaming is
efficient on a characteristic scale called the Jeans length,
corresponding roughly to the distance on which neutrinos can travel in
a Hubble time. For ultra-relativistic neutrinos, the Jeans length is
by definition equal to the Hubble radius $c/H$, but for
non-relativistic ones it grows at a slower rate than $c/H$ (in
comoving coordinates, it even decreases with time during matter
domination). Neutrinos with masses smaller than approximately
$T_{\rm dec} \sim 0.3$ eV are still relativistic at the
time of last scattering, and their direct effect on the CMB
perturbations is identical to that of massless neutrinos. For bigger
masses, the decrease of the free-streaming scale is felt by
perturbations which enter inside the Hubble radius before decoupling,
which results in a small enhancement of the acoustic peaks with
respect to the massless situation.

In the intermediate mass range from $10^{-3}$ eV to $0.3$ eV, the
transition to the non-relativistic regime takes place during structure
formation, and the matter power spectrum will be directly affected in
a mass-dependent way. Wavelengths smaller than the current value of
the neutrino Jeans length are suppressed by free-streaming. The
largest observable wavelengths, which remain always larger than the
neutrino Jeans length, are not affected.  Finally, there is a range of
intermediate wavelengths which become smaller than the neutrino Jeans
length for some time, and then encompass it again: these scales
smoothly interpolate between the two regimes.  The net signature in
the matter power spectrum is a damping of all wavelengths smaller than
the Hubble scale at the time of the transition of neutrinos to a
non-relativistic regime \cite{Hu:1997mj}
\begin{equation}
k > k_{\rm nr} = 0.026 \left(\frac{m_\nu \, \Omega_m}
{1 ~{\rm eV}}\right)^{1/2} 
h~{\rm Mpc}^{-1}.
\end{equation}
where $\Omega_m$ is the contribution of matter to the critical
density. The damping is maximal for wavenumbers bigger than the
current free-streaming wavenumber $k_{\rm FS}$
\begin{equation}
k > k_{\rm FS} = 0.63 \left(\frac{m_\nu}{1 ~{\rm eV}}\right) h~{\rm
Mpc}^{-1}.
\end{equation}

We have summarized both the background and the direct effects of the
neutrino masses on the CMB and LSS perturbations. The total signature
is difficult to describe analytically. However, one should remember
that for masses of order $1$ eV or less, the dominant effect is the
one induced by free-streaming on the matter power spectrum.
Therefore, the usual strategy is to combine CMB and LSS measurements,
where the former roughly fix most of the cosmological parameters,
while the latter is sensitive to $k_{\rm FS}$ and provides bounds on
the neutrino mass.

\subsection{Combined effects}

In a situation with $N$ thermalized massive neutrino species, the
cosmological model should include an equal number of parameters,
namely $(m_1, \ldots, m_N)$.  However, at first order such a model could
be described by only two parameters, $N$ and the sum of all individual
masses $\sum_{i=1,N} m_i$. This choice not only simplifies the
problem, but also provides the correct contribution of neutrinos to
the total energy density both at early times, when all neutrinos are
ultra-relativistic and the radiation density depends only on $N$, and
at a late epoch when at least one neutrino mass is large compared with
the temperature and the density is given in good approximation by the
total mass.
 
However, for a precise description $N$ and $\sum_{i=1,N} m_i$ are not
the only relevant parameters. As an example, let us compare two
scenarios with three neutrinos but different mass spectra: a
degenerate case with $(m_0, m_0, m_0)$ and a case with $(3m_0,
\epsilon, \epsilon)$, where $\epsilon \ll m_0$.  In both scenarios the
neutrino density is the same for early and late stages of the
Universe.  But at intermediate temperatures of the order $m_0 \alt
T_{\nu} \alt 3 \, m_0$, the energy densities are different.  It is easy
to show that the expansion rate is temporarily enhanced in the second
scenario, but this will only have a small signature in the CMB
spectrum for $m_0 \agt 0.2$ eV. On the other hand, the free-streaming
wavenumber $k_{\rm FS}$ of the heaviest neutrino is larger in the
second scenario. Thus, in principle we expect more damping and sharper
bounds on the mass in the degenerate case.
 
Nowadays, after many years of experimental effort, we know that
neutrinos must be massive in order to explain the evidences for flavor
oscillations from measurements of atmospheric and solar neutrinos,
independently confirmed by data from the detection of neutrinos
from artificial sources at experiments such as K2K and Kamland.
These results lead to specific differences between the individual
neutrino masses, at two different scales: $\Delta m^2_{\rm atm}\simeq
2.5 \times 10^{-3}$ eV$^2$ and $\Delta m^2_{\rm sun}\simeq 7 \times
10^{-5}$ eV$^2$ \cite{Maltoni:2004ex}. But one of the masses remains
unconstrained, which reflects the fact that oscillation experiments
can not fix the absolute scale of neutrino masses. It is clear that,
considering the present cosmological data, in order to have a
measurable effect the three neutrinos should have roughly the same
mass, following a degenerate scheme $(m_0, m_0, m_0)$, where $m_0 \agt
0.2$ eV.
 
On the other hand, the positive results from the LSND experiment point
to neutrino oscillations with $\Delta m^2 \sim {\cal O}(1~{\rm eV}^2)$.
The less disfavored scenario that could explain the LSND data,
together with the results of atmospheric and solar neutrino
experiments, contains 4 neutrinos following a 3+1 scheme where one of
them is much heavier than the others (see for instance
\cite{Schwetz:2003pv}) and the fourth neutrino must be sterile.
It has been shown \cite{DiBari:2001ua,Abazajian:2002bj} that
all four neutrino models lead to a full thermalization of the sterile
neutrino before BBN, so they are disfavored by BBN. However, current CMB
and LSS data can not completely rule out this possibility
\cite{Hannestad:2003xv,Hannestad:2003ye}.

The most general scenario is that of a cosmological model with $N^{\rm
  nr}$ thermally-decoupled massive neutrinos and extra relativistic
degrees of freedom, parametrized by $N^{\rm r}$ not necessarily
integer.  This model is described by a set of $N^{\rm nr}+1$
parameters: $(m_1, \ldots, m_{N^{\rm nr}}, N^{\rm r})$. However, since
such a parameter space is too large for a systematic analysis we will
consider restricted cases described by only two parameters: the total
effective neutrino number $N_{\rm eff}=N^{\rm r}+N^{\rm nr}$ and the
total mass $M$. In order to check the impact of the distribution of
the total neutrino mass among the individual states, we will study two
cases:
\begin{itemize}
\item the model that we call {\it degenerate} has $N_{\rm eff}$
  massive neutrinos with the same mass. Let us emphasize that this is
  a simplified model where the physical interpretation of non-integer
  values of $N_{\rm eff}$ is not obvious.
\item a second model the we denote {\it 1+r} has only one neutrino
  with a significant mass, while the other $N_{\rm eff}-1$ species are
  ultra-relativistic.
\end{itemize}
These two models are not chosen arbitrarily. First, the {\it
degenerate} model includes the standard situation with only 3 flavor
neutrinos degenerate in mass, while the {\it 1+r} model includes the
3+1 scenario described above. Second, and most importantly, the {\it
degenerate} and {\it 1+r} models appear as limiting situations of the general
case once the parameters $N_{\rm eff}$ and $M$ have been
fixed. Indeed, the former has the smallest possible value of the 
free-streaming wavenumber, while the latter has the biggest $k_{\rm FS}$.
For any intermediate model (like, for instance,
the third case studied in \cite{Hannestad:2003ye}, with three
effectively massless standard neutrinos and $N_{\rm eff}-3$ species
with equal mass), the observational bounds deduced from the CMB and LSS
observations should lay between those that we get in the
{\it degenerate} and {\it 1+r} limits.

\section{Method and data used}
\label{method_data}

The WMAP spectrum and many other cosmological data can be accurately
fitted with a six-parameter flat $\Lambda$CDM model
\cite{Spergel:2003cb,Tegmark:2003ud}, described by the Hubble
parameter $h$, the fractional density of matter $\Omega_m =
1-\Omega_{\Lambda}$, the baryon density in dimensionless units
$\omega_b = \Omega_b h^2$, the optical depth to reionization $\tau$,
and finally, the amplitude and the spectral tilt of primordial
perturbations $(A_s, n_s)$. Most of our calculations correspond to a
model with eight parameters: the six previous ones plus $N_{\rm eff}$
and $M$, as previously defined for the {\it degenerate} and {\it 1+r}
models. We will also study the consequences of the presence of a
background of primordial gravitational waves, which would contribute
to the CMB anisotropy spectrum. For this case, our parameter space
will be ten-dimensional, adding the tensor-to-scalar ratio $r$ and the
tensor tilt $n_t$. In all cases, we use the CMBFAST code
\cite{Seljak:1996is} to calculate the power spectra.

In order to compare the theoretical cosmological models with current
observations, we use a Bayesian grid-based method described in some
previous works (e.g.\ \cite{Crotty:2003th}), instead of the widely
used Monte Carlo Markov Chains (MCMC) technique employed for instance
in \cite{Tegmark:2003ud,Verde:2003ey}. The former has the inconvenient
of being considerably slower from the computational point of view, and
the advantage of being very robust if the hypersurfaces with the same
likelihood in parameter space have very complicated shapes (for
instance, an elongated and curved {\it banana} shape).  Also, the MCMC
method cannot deal with likelihood distributions with nearly
degenerate local minima; mathematically, the existence of local minima
cannot be disproved, but in practice, using current data and the wide
variety of models discussed in the literature, such a situation never
appeared.  In order to analyze unusual models with many parameters, it
is safe to use at least once a grid-based method, avoiding surprises
that could arise from parameter degeneracies. However, our analysis
always shows quasi-ellipsoidal likelihood contours for various
combinations of parameters. Thus we believe that a MCMC method would
give similar results.

For each combination of data set and priors, we plot the Bayesian
likelihood of each point in the two-dimensional space of neutrino
parameters $(M, N_{\rm eff})$ after marginalization over the other
free parameters. Actually, for simplicity we approximate the
integration process required for a true marginalization by a
maximization of the likelihood over the remaining parameter space.
Integration and maximization are known to be equivalent in the case of
a multigaussian likelihood distribution and we expect the maximization
technique to be reasonably accurate in our case, where we obtained
quasi-ellipsoidal contours. We checked this explicitly by computing
the marginalized likelihood of each of the six $\Lambda$CDM
parameters: our results are very close to those of
\cite{Spergel:2003cb,Tegmark:2003ud}, obtained with a MCMC technique
and marginalized by integration over the likelihood.  Of course, our
maximization routine needs to compute the likelihood not only at grid
points, but also in between. At a given point in parameter space, this
is done by interpolating quadratically each value of the $C_l$ and
$P(k)$ from the nearest neighbors of the grid.

Our CMB data include the 1348 correlated points of the Wilkinson
Microwave Anisotropy Probe (WMAP), which measure the temperature
$\times$ temperature (TT) \cite{Hinshaw:2003ex} and temperature
$\times$ E-polarization (TE) \cite{Kogut:2003et} correlation functions
on the CMB sky. WMAP provides the best available data on multipoles $l
\leq 900$. For constraining the TT spectrum on smaller scales, we
employ the results of the Arcminute Cosmology Bolometer Array Receiver
(ACBAR) experiment \cite{Kuo:2002ua,Goldstein:2002gf}. We remove the
highest band powers (probing the region $l>1800$) which could be
contaminated by foregrounds. The published ACBAR band powers are
decorrelated, so it is self-consistent to use only the first 11 points
of data.

Our LSS data set consists of 32 correlated points from the 2 degree
Field (2dF) galaxy redshift survey \cite{Percival:2001hw}, covering
the range $0.02 \, h~{\rm Mpc}^{-1} < k < 0.15 \, h~{\rm Mpc}^{-1}$, and of
19 decorrelated points from the Sloan Digital Sky Survey (SDSS)
\cite{Tegmark:2003uf} in the range $0.015 \, h~{\rm Mpc}^{-1} < k < 0.20
\, h~{\rm Mpc}^{-1}$. In order to compare the smallest wavelengths with
the data, it is necessary to take into account small deviations with
respect to the linear power spectrum.  Following the analysis in
\cite{Tegmark:2003ud}, we compute non-linear corrections for each
point in parameter space using the numerical procedure described in
Appendix C of \cite{Smith:2002dz}. Since these corrections rely on
N-body simulations of a pure CDM Universe they might not be perfectly
optimized for hot plus cold dark matter.  However, the models studied
here include only a small fraction of non-relativistic neutrino
density. Therefore, at first order the fitting formula of
\cite{Smith:2002dz} should be reasonably accurate in our case (and
certainly better than introducing no corrections at all). Each
redshift survey is expected to constrain the total matter power
spectrum modulo a global normalization factor called the bias $b$.
Unless otherwise stated, all our results were obtained after
marginalizing over the 2dF and SDSS bias, treated like two free
parameters.

We will also use a prior on the current value of the Hubble parameter,
measured by the Hubble Space Telescope (HST) Key project
\cite{Freedman:2000cf}: $h = 0.72 \pm 0.08$ (1$\sigma$). Finally, we
will impose some constraints on the current density of the
cosmological constant deduced from the redshift dependence
of type Ia supernovae luminosity.  For a flat universe,
Perlmutter et al.\ \cite{Perlmutter:1998np} give a conservative bound
$\Omega_m = 0.28 \pm 0.14$ (1$\sigma$) that we denote as the SN99
prior. In addition, we will calculate the impact of using the
more restrictive result $\Omega_m = 0.28 \pm 0.05$ (1$\sigma$) from the
recent work by Tonry et al.\ \cite{Tonry:2003zg}, that we label the
SN03 prior.

\section{Results and comparison with previous works}
\label{results}

\subsection{Degeneracies and priors}

In order to understand the impact of each data set and prior on the
final results, we will introduce them step by step in the calculation
of the likelihood. We will first focus on the {\it degenerate} model.

\begin{figure}
\includegraphics[width=.8\textwidth]{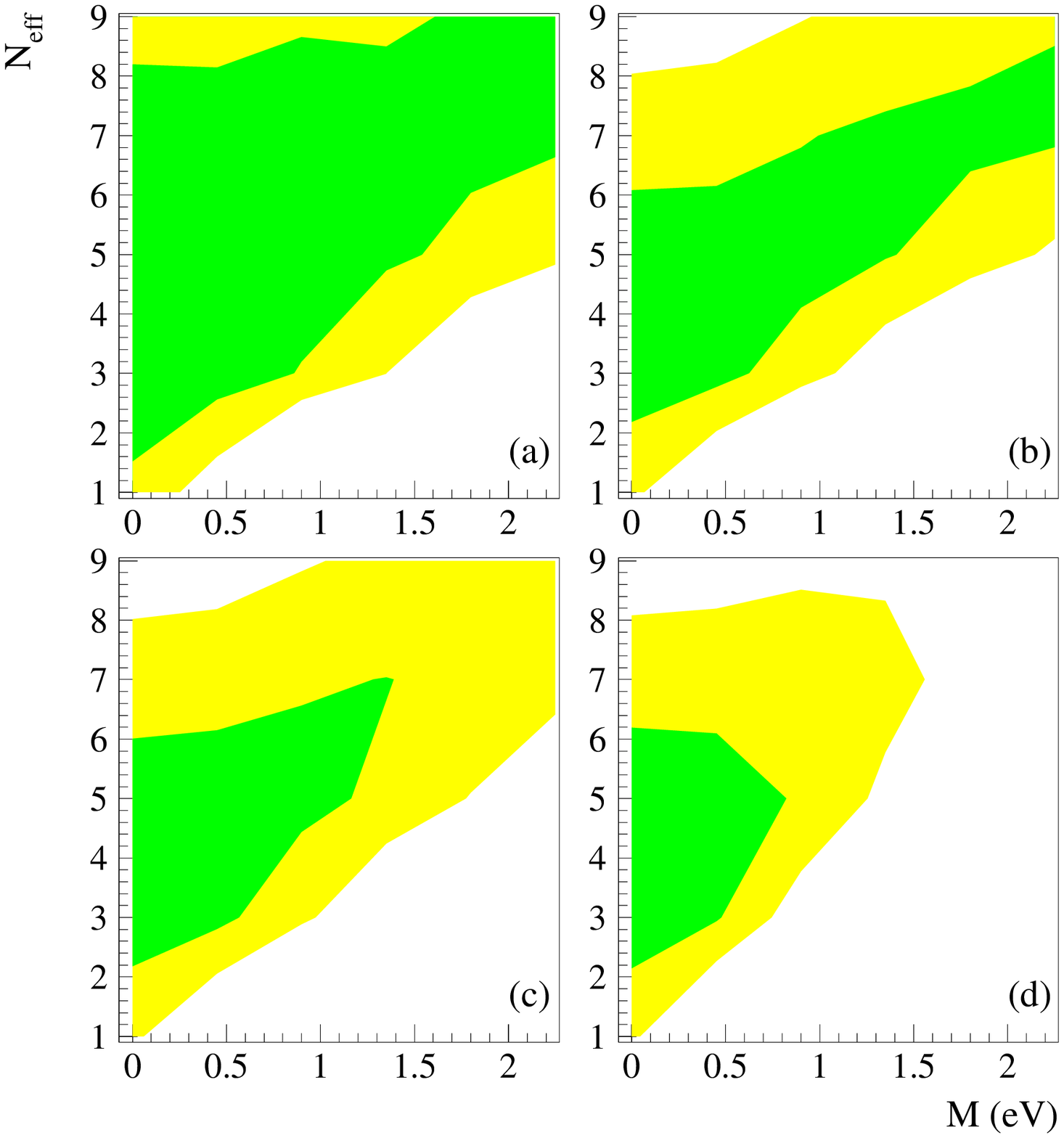}
\caption{\label{fig1abcd} 
  Two-dimensional likelihood in $(N_{\rm eff}, M)$ space, marginalized
  over the six remaining parameters of the model. We plot the
  1$\sigma$ (green / dark) and 2$\sigma$ (yellow / light) allowed
  regions. Here we used CMB (WMAP \&
  ACBAR) and LSS (2dF \& SDSS) data, adding different external priors
  as defined in section \ref{method_data}: (a) no priors, (b) HST, (c)
  HST+SN99, (d) HST+SN03.}
\end{figure}

We start using only CMB and LSS data without any prior, apart from the
top-hat priors defined implicitly by the boundaries of our grid. We
look for the two-dimensional probability distribution of the neutrino
parameters in the range $0 < M < 2.25 \, {\rm eV}$ and $1 < N_{\rm
  eff} < 9$. We find that most of this parameter space is allowed at
$2\sigma$, as shown in panel (a) of Fig.\ \ref{fig1abcd}. Actually
some of the grid boundaries are reached by the allowed models
\footnote{For fixed $N_{\rm eff} <2$ or $N_{\rm eff} > 8$, the
  best-fit models reach the extreme values of $h \in [0.58, 0.90]$ and
  $\omega_m \in [0.11, 0.25]$. However, the global maximum-likelihood
  fit has $(M, N_{\rm eff}) = (0, 3)$ and values of the other
  cosmological parameters well inside the grid boundaries.}, unlike in
the rest of our analysis where they have no influence on the $2\sigma$
allowed regions.

The physical explanation for such loose constraints is well-known: 
when the density of relativistic relics is left free, there is a
parameter degeneracy between 
\begin{equation}
\omega_r \equiv \Omega_r h^2 
= 2.47 \times 10^{-5} \, h^2 
\left(\frac{T_{\rm CMB}}{2.725 \, {\rm K}}\right)^4
\left[ 1 + \frac{7}{8} \left( \frac{4}{11} \right)^{4/3} 
N_{\rm eff} \right] 
\end{equation}
and $\omega_m \equiv \Omega_m h^2$. Indeed, one can vary these two
quantities in the same proportion while keeping fixed $z_{\rm eq}$,
the redshift of equality between matter and radiation. In order to
remove this degeneracy, it is necessary to impose some priors on $h$
and $\Omega_m$. Thus we repeat the same analysis, using now the HST
prior. In this case, we obtain a band of allowed models shown in panel
(b) of Fig.~\ref{fig1abcd}. This region stretches up to the maximal
values of $M$ and $N_{\rm eff}$ in our grid, so we cannot derive yet
limits on these parameters. However, the other grid bounds are now
irrelevant, since none of the models allowed at $2\sigma$ ever reaches
them. The $1\sigma$ preferred region includes models with ($M =0$,
$N_{\rm eff}=3.04$), showing that there is no evidence for massive
neutrinos and/or extra relativistic degrees of freedom. However, it is
interesting to see that large departures from the standard
$\Lambda$CDM model cannot be excluded.

These results show clearly how difficult is to put bounds on the
neutrino mass in presence of an excess of relativistic relics during
RD, and vice versa. For instance, if we assume $N_{\rm eff}=3$, we get
a $2\sigma$ upper bound $M < 0.8$ eV, while for $N_{\rm eff}=6$ this
bound spectacularly increases to $M < 2.2$ eV.  This trend was already
observed in
\cite{Lesgourgues:2001he,Elgaroy:2003yh,Hannestad:2003xv,Hannestad:2003ye},
and can be explained as follows. Suppose that we start from the
best-fit standard model with ($M =0$, $N_{\rm eff}=3$), and that we
increase $N_{\rm eff}$ in such way that the radiation density in the
early Universe is multiplied by a factor $\alpha$. Then, in order to
keep the CMB power spectra roughly constant, we should maintain a
fixed value both for $z_{\rm eq}$ and $\Omega_{\Lambda}$. This would
be achieved with a transformation of the type $(\omega_m, h)
\longrightarrow (\alpha \, \omega_m, \sqrt{\alpha} \, h)$. However, it
is well-known that the shape of the matter power spectrum is given
roughly by the parameter $\Gamma = \Omega_m h$, that scales like
$\sqrt{\alpha} \, \Gamma$ with the previous transformation, meaning
that the spectrum will have more power on small scales relatively to
large scales. A neutrino mass can balance this increase through the
free-streaming effect. Therefore, models with high values of $N_{\rm
  eff}$ are compatible with larger neutrino masses.

\begin{figure}
\includegraphics[angle=-90,width=.48\textwidth]{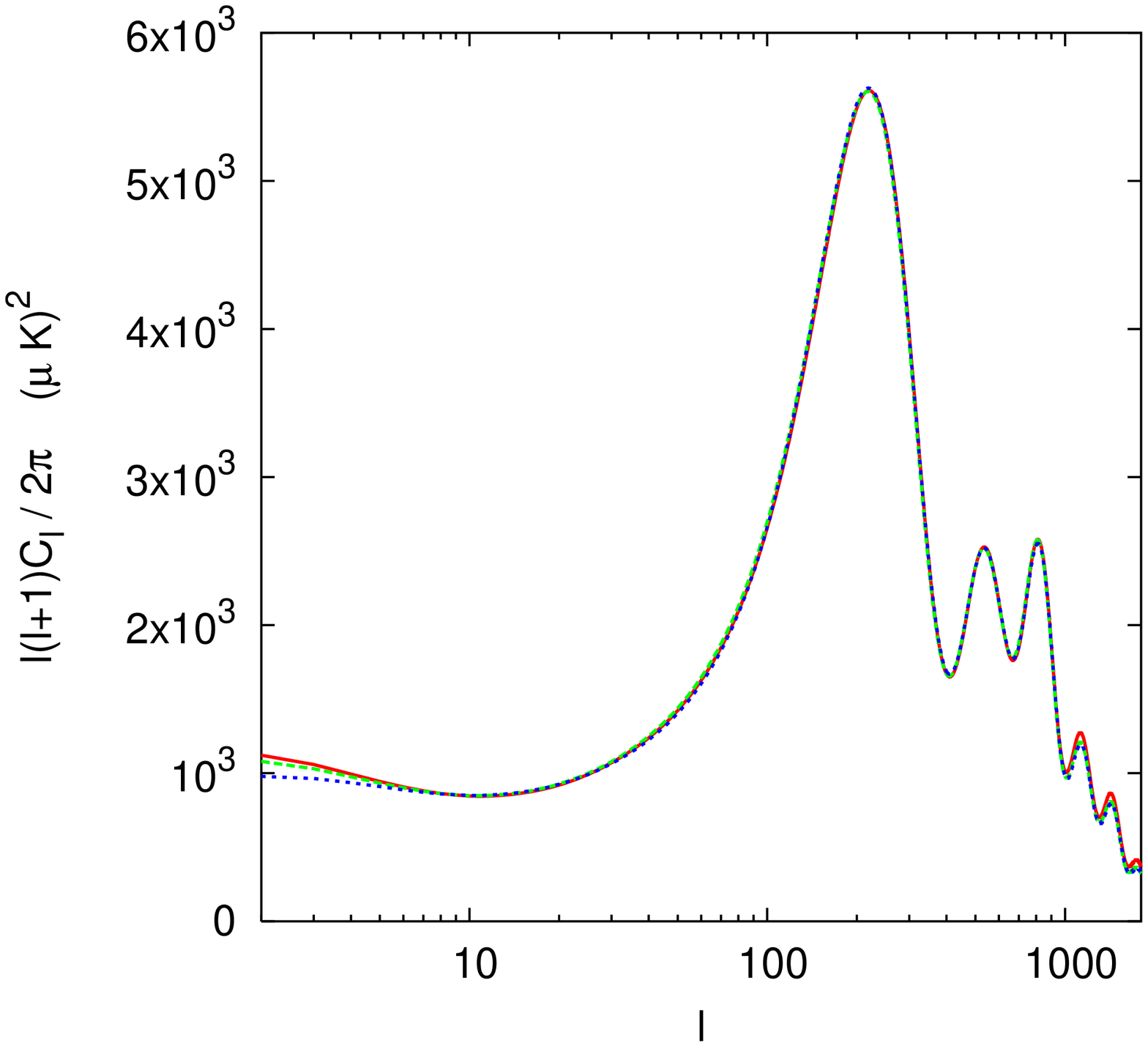}
\includegraphics[angle=-90,width=.48\textwidth]{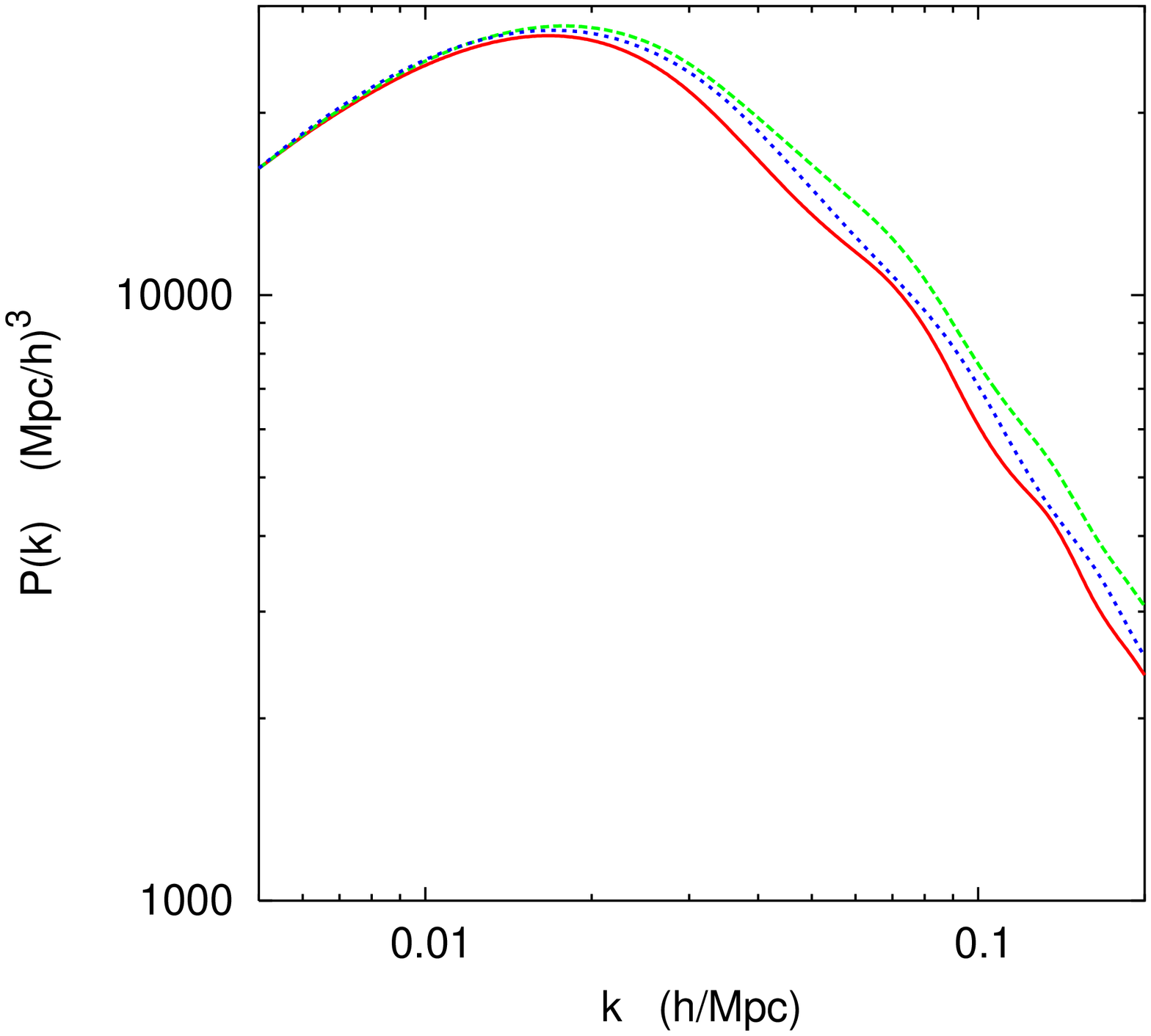}
\caption{\label{figdeg} Illustration of the main parameter degeneracy
affecting our results. For three particular models, we plot on the
left the CMB temperature spectrum (normalized to WMAP) and on the
right the matter power spectrum (two of them have been rescaled by
hand for clarity). See the main text for details.}
\end{figure}

We illustrate this parameter degeneracy in Fig.\ \ref{figdeg}, where
we show the CMB temperature spectrum (normalized to WMAP) and the
matter power spectrum for three particular models.  The first model
(red/solid curves) is the best-fit model for the standard case $(M,
N_{\rm eff})=(0,3)$. In the second model (green/dashed curves), we
have increased the number of degrees of freedom to $N_{\rm eff}=7$ and
performed the transformation on ($\omega_m$, $h$) as previously
discussed, leaving both $z_{eq}$ and $\Omega_M$ invariant. We have
also increased $n_s$ a little bit. On the CMB figure it is difficult
to distinguish the curves because they perfectly overlap, especially
for the best measured scales (those of the first acoustic peak), while
the difference at $l<10$ is masked by cosmic variance. But as
expected, the shape of the matter power spectrum is modified, with
more power on small scales relatively to large scales. This leaves
plenty of room for models with a significant neutrino mass and
free-streaming effect, like the one featured here (blue/dotted curve)
which has $(M, N_{\rm eff})=(2.25 \, {\rm eV},7)$.

In order to obtain bounds on the mass, it is necessary to add more
restrictive priors. In particular, the supernova priors should help to
better constrain $\Omega_m$ (around the central value $0.28$) and
therefore also the shape parameter $\Gamma$ (around $\Omega_m h = 0.28
\times 0.72 = 0.20$). This value of $\Gamma$ is in very good agreement
with the constraint obtained from LSS data, assuming no neutrino mass:
the SDSS power spectrum points to $\Gamma = 0.21 \pm 0.03$
\cite{Tegmark:2003uf}. Therefore, we expect the SN prior to restrict
the possibility of a large neutrino free-streaming effect and to
improve the upper bound on $M$. This is what we observe by
successively adding the SN99 and SN03 priors in our analysis. The
corresponding contours, shown in panels (c) and (d) of Fig.\ 
\ref{fig1abcd}, are the main results of this work. With the SN03
prior, our one-dimensional bounds on the effective neutrino number
(marginalized over $M$) read $1.6 < N_{\rm eff} < 7.2$ at $2\sigma$,
in very good agreement with our previous results $1.4 < N_{\rm eff} <
6.8$ \cite{Crotty:2003th} obtained without SDSS, no SN priors and
fixing $M=0$.  The one-dimensional upper bound on the total neutrino
mass (marginalized over $N_{\rm eff}$) is $M < 1.1$ eV at $2\sigma$.
For comparison, we list the bounds for fixed integer values of $N_{\rm
  eff}$ are in Table \ref{table1}, for different priors.  Since we use
more recent 2dF and SDSS data, as well as more restrictive SN priors,
we are not surprised to find stronger bounds than Hannestad \& Raffelt
\cite{Hannestad:2003ye}.

\begin{table}
\begin{ruledtabular}
\begin{tabular}{ccccc}
& \multicolumn{2}{c}{degenerate} & \multicolumn{2}{c}{1+r} \\
$N_{\rm eff}$ 
& no priors & HST+SN03 & no priors & HST+SN03 \\
\hline
3 & 1.0 & 0.6 & 0.8 & 0.6 \\
\hline
4 & 1.5 & 0.8 & 1.2 & 0.8 \\
\hline
5 & 2.0 & 1.0 & 1.6 & 1.1 \\
\hline
6 & --  & 1.1 & 1.9 & 1.4 \\
\hline
7 & --  & 1.0 & -- & 1.5 \\
\end{tabular}
\end{ruledtabular}
\caption{\label{table1}The 2$\sigma$ upper bound on the total
neutrino mass $M$ (eV), 
after marginalization over the six
cosmological parameters of the flat $\Lambda$CDM model, for
particular values of $N_{\rm eff}$. We show the results for two
limiting cases of splitting the total mass between the neutrino states:
{\it degenerate} (all neutrinos with the same mass) and
{\it 1+r} (one massive neutrino and the other treated as relativistic relics).
Here we have used the full CMB and LSS data set, either alone
(no priors) or combined with the HST and SN03 priors. For large
$N_{\rm eff}$ values and in absence of priors,  
the upper bound is larger than the maximal value
of $M$ in our grid (2.25 eV).}
\end{table}

\subsection{Role of LSS data}

\begin{figure}
\includegraphics[width=.8\textwidth]{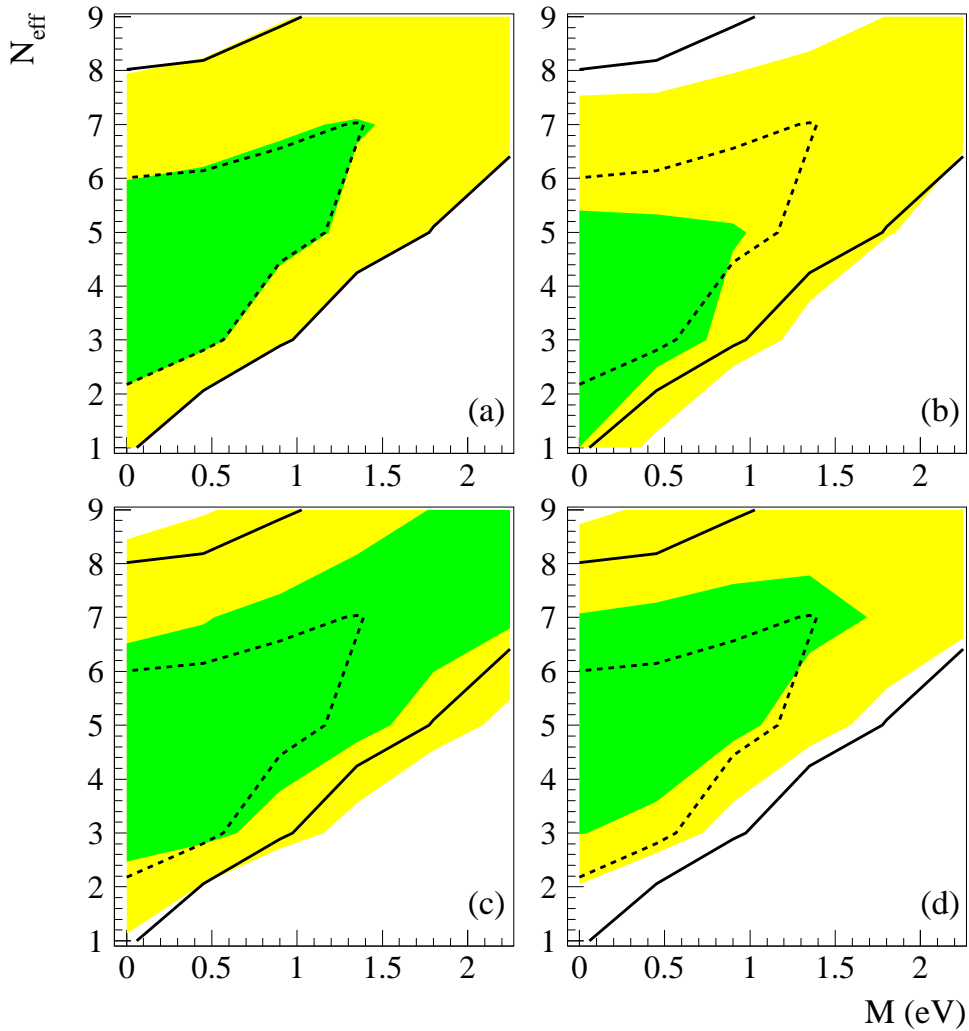}
\caption{\label{fig3abcd} Impact of various assumptions related to the
LSS data set. The default model for which we show 1$\sigma$
(dashed) and 2$\sigma$ (solid) contours was obtained with WMAP, ACBAR,
2dF, SDSS, plus the HST and SN99 priors. Each panel shows the
1$\sigma$ (green / dark) and 2$\sigma$ (yellow / light) allowed
regions (a) when including the 2dF bias prior \cite{Peacock:2001gs},
(b) without the SDSS data, (c) without the 2dF data, (d) without
non-linear corrections.  }
\end{figure}

So far we have used the LSS data as an indication of the shape of the
matter power spectrum, but not its overall amplitude. This amplitude
is difficult to measure, because of possible differences between the
two-point correlation function of luminous galaxies and that of
matter, a problem known as the bias uncertainty. The 2dF team has
established that the bias $b$ is almost scale-independent, and derived
some constraints either on the redshift distortion parameter $\beta =
\Omega_m^{0.6} / b$ \cite{Peacock:2001gs} or directly on $b$
\cite{Verde:2001sf}. These two results must be employed with great
care since the bias is expected to depend on the mean luminosity and
redshift of each particular galaxy sample. In order to use a
self-consistent bias prior, we would need to compute some correction
factors for each model (see for instance refs.\ 
\cite{Lahav:2001sg,Verde:2003ey,Elgaroy:2003yh}).  This technically
difficult procedure, that relies on many assumptions, is beyond the
goal of the present paper and we prefer to conservatively discard any
bias prior, as in \cite{Tegmark:2003ud}. Just for indication, we tried
to repeat the previous analysis with a very naive bias prior.  Instead
of leaving the 2dF bias as a free parameter, we tried to add the
constraint $\Omega_m^{0.6}/b_{\rm 2dF} = 0.43 \pm 0.07$
\cite{Peacock:2001gs} to our full set of data and priors (which
includes all the CMB+LSS data, the HST prior, and one of our two
supernovae priors). As shown in panel (a) of Fig.\ \ref{fig3abcd}, our
results remain unchanged. This is consistent with the analysis of
Elgar\o y \& Lahav \cite{Elgaroy:2003yh}, who treat the bias prior in a
detailed way and find no impact on the neutrino mass determination.

Since the LSS data plays a crucial role in constraining the neutrino
mass, it is worth comparing the impact of the 2dF and SDSS power
spectra. We go back to a data set consisting of CMB+LSS+HST+SN99, and
remove either the SDSS or 2dF spectrum from the analysis.  The
results, shown in panels (b) and (c) of Fig.\ \ref{fig3abcd}, should
be compared with the combined analysis previously shown in (c) of
Fig.\ \ref{fig1abcd}. The SDSS power spectrum appears to be much more
conservative, in good agreement with previous papers: the WMAP+SDSS
constraint on the neutrino mass for $N_{\rm eff}=3$ is as large as
$M<1.74$ eV \cite{Tegmark:2003ud}, while a WMAP+other CMB+2dF analysis
gives $M<0.69$ eV \cite{Spergel:2003cb}.  Consistently, our combined
analysis gives intermediate results: for $N_{\rm eff}=3$ our
WMAP+ACBAR+2dF+SDSS bound is $M < 0.9$ eV. Our results seem also
consistent with the recent analysis in \cite{Barger:2003vs}, where the
corresponding bound $M<0.75$ eV was found including data from both
galaxy surveys up to $k \alt 0.15 \, h~{\rm Mpc}^{-1}$.

\begin{figure}
\includegraphics[angle=-90,width=.48\textwidth]{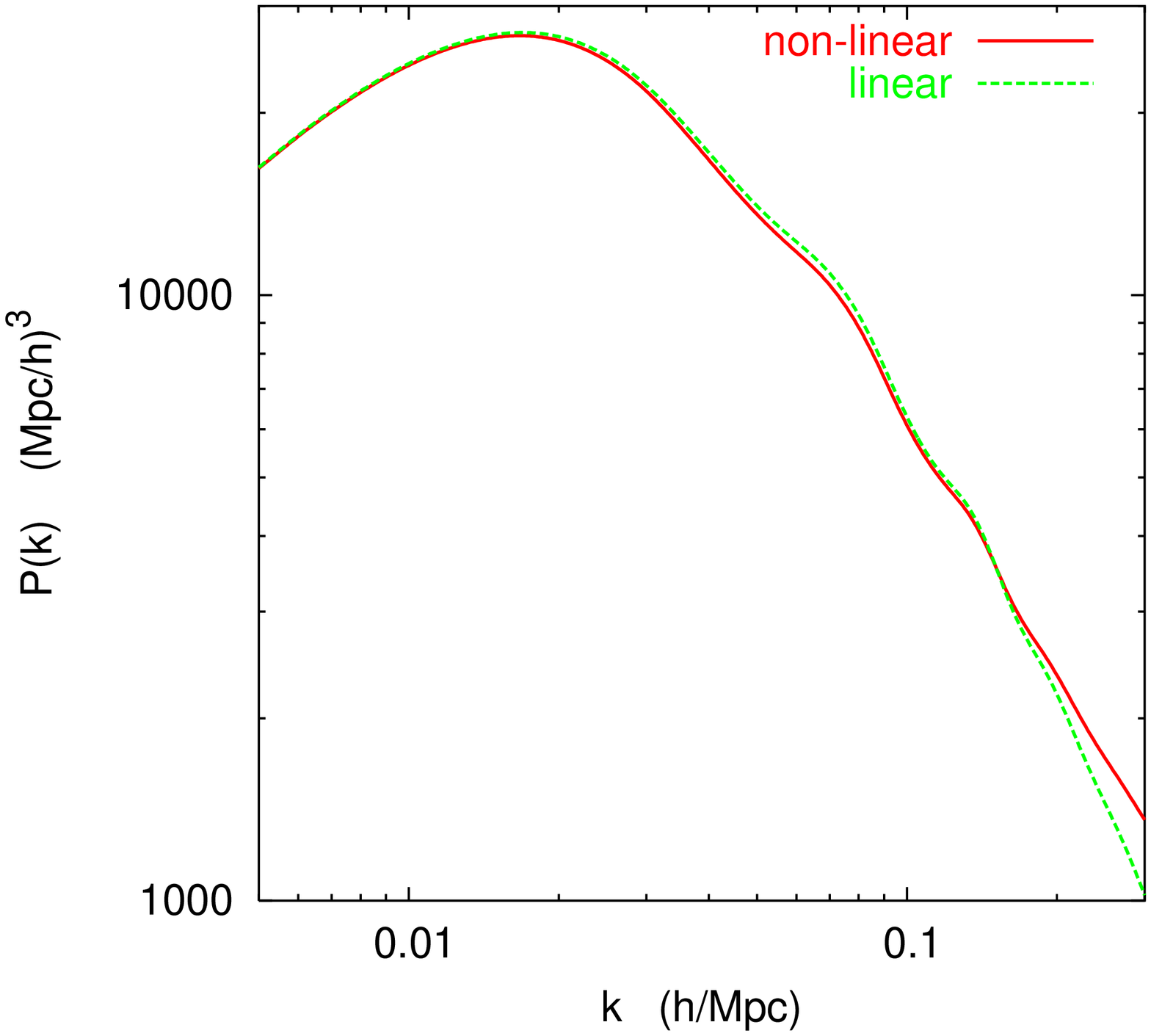}
\caption{\label{fig4}Matter power spectrum for the best-fit model with 
  $(M, N_{\rm eff})=(0,3)$ , plotted with and without non-linear
  corrections. The difference becomes significant in the region $k >
  0.15 \, h~{\rm Mpc}^{-1}$ probed by the last two points in the SDSS
  data set.}
\end{figure}

Note that in order to employ the SDSS data until $k_{\rm
  max} \simeq 0.20 \, h~{\rm Mpc}^{-1}$, it is crucial to include the
non-linear corrections to the matter power spectrum, in particular for
the $N_{\rm eff}$ bounds. The panel (d) of Fig.\ \ref{fig1abcd} was
obtained with all the CMB and LSS data, plus the HST and SN99 priors,
but in absence of non-linear corrections. A comparison with Fig.\ 
\ref{fig1abcd}c shows that the constraints on $N_{\rm eff}$ are lifted
by one unit.  This can be easily understood. Fig.\ \ref{fig4} shows a
typical power spectrum with and without non-linear corrections. The
linear power spectrum has less power on small scales, i.e.\ a smaller
effective shape parameter $\Gamma$. As explained earlier in this
section, this can be easily compensated by an appropriate increase in
$N_{\rm eff}$, $\omega_m$ and $h$, while leaving the CMB spectrum
almost invariant.

\subsection{Adding tensors}

We have discussed so far the dependence of our results under the
choice of data set and priors. However, generally speaking, the bounds
on a particular parameter also depend on assumptions concerning the
underlying cosmological model, a simple six-parameter flat
$\Lambda$CDM model in our case. It is clear that by adding extra
physical ingredients that would compensate the effect of neutrinos, we
would relax the bounds on ($N_{\rm eff}$, $M$). It is inviable to
perform a systematic study of all the $\Lambda$CDM variants proposed
in the literature, but one of them deserves a particular interest.
Indeed, the six-parameter $\Lambda$CDM relies on the existence of
super-horizon cosmological fluctuations at early times, which strongly
suggests that perturbations are of inflationary origin.  But inflation
also predicts a background of primordial tensor perturbations: the
question is whether these gravitational waves are large enough to
contribute to large-scale CMB anisotropies (fundamentally, this
depends on the energy scale of inflation). It is thus important to see
how the neutrino parameter bounds evolve in presence of two extra
parameters, the relative amplitude and tilt of the primordial tensor spectrum
$(r, n_t)$.

\begin{figure}
\includegraphics[width=.5\textwidth]{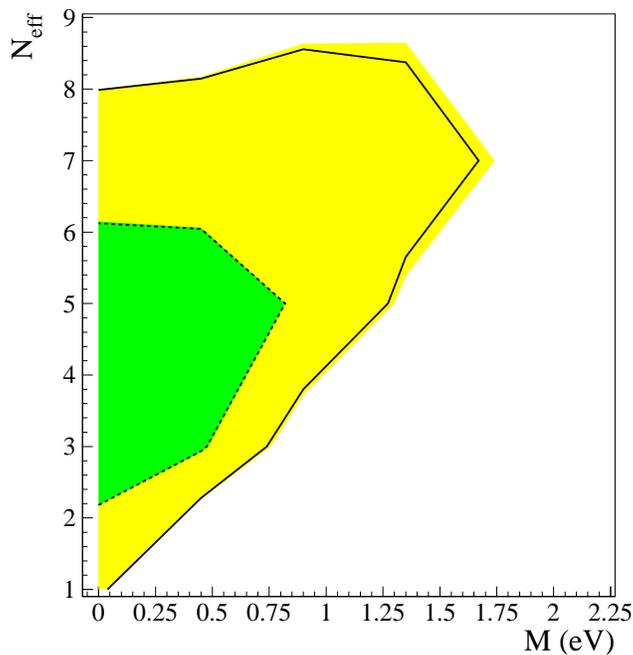}
\caption{\label{fig_tensor} 1$\sigma$ (green / dark) and 2$\sigma$
(yellow / light) allowed regions in ($M$, $N_{\rm eff}$) space,
marginalized over the eight other free parameters of the flat
$\Lambda$CDM+tensors model.  For comparison, we show the contours
corresponding to the case with no gravitational waves. The difference
is very small, showing the ability of the data to make a clear
difference between the effect of neutrinos and gravitational waves.  }
\end{figure}

Previous analyses \cite{Spergel:2003cb,Tegmark:2003ud} showed that for
eight-parameter models (flat $\Lambda$CDM + tensors) a significant
contribution of gravitational waves is disfavored. However, one could
expect that in a ten-dimensional model (flat $\Lambda$CDM + tensors,
with $N_{\rm eff}$ and $M$), a new parameter degeneracy would show up
and relax the various bounds.  We performed such an analysis for our
full CMB and LSS data set, adding the HST and SN03 priors. The
resulting two-dimensional likelihood for ($N_{\rm eff}$, $M$),
marginalized over the other eight parameters, is shown in Fig.\ 
\ref{fig_tensor}. It is almost indistinguishable from that with
a vanishing contribution of gravitational waves, which shows that the
current data is clearly able to distinguish between the respective
effects of tensors and neutrinos. Thus the cosmological bound on
neutrino masses is robust with respect to tensors. This robustness
also holds when including a non-adiabatic, incoherent contribution to
the power spectrum such as those predicted by topological defects, as
shown in a very recent work \cite{Brandenberger:2004kc}.

\subsection{Impact of mass splitting}

Our aim is to constrain cosmological models with an arbitrary number
of massive neutrinos and with extra relativistic degrees of
freedom. However, so far we reduced the analysis to two parameters
($M$, $N_{\rm eff}$), with the implicit assumption that all neutrinos
were degenerate in mass. We have not discussed the fact that for a
fixed total number of degrees of freedom $N_{\rm eff}$ and total mass
$M$, the evolution of cosmological perturbations depends on the
splitting of the mass between the different species.

\begin{figure}
\includegraphics[width=.5\textwidth]{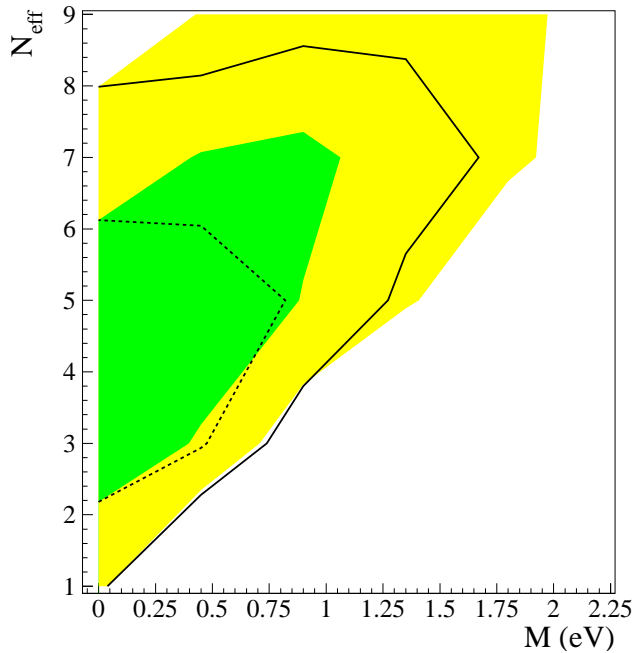}
\caption{\label{fig_split} 1$\sigma$ (green / dark) and 2$\sigma$
(yellow / light) allowed regions in ($M$, $N_{\rm eff}$) space for the
{\it 1+r} model, compared with the {\it degenerate} model (dashed
1$\sigma$ and solid 2$\sigma$ contours).  Here we have included all
CMB and LSS data, plus the HST and SN03 priors.}
\end{figure}

As explained in section \ref{effects}, for fixed ($M$, $N_{\rm eff}$),
in some sense the opposite case to the {\it degenerate} model is the
{\it 1+r} scenario, for which all the mass corresponds to a single
neutrino eigenstate, instead of being equally shared.  We built a
second grid of {\it 1+r} models and analyzed it with our most
restrictive set of data (WMAP+ACBAR+2dF+SDSS) and priors
(HST+SN03). We show in Fig.\ \ref{fig_split} the new allowed regions,
compared with the previous ones.
In the limit of small mass and small $N_{\rm eff}$, the CMB
and LSS power spectra are almost identical in the two cases, because
the effect of the mass splitting is second-order with respect to that
of the total mass. A priori, this does not guarantee that the ($M$,
$N_{\rm eff}$) iso-likelihood contours are asymptotically equal,
because we are doing a Bayesian analysis and the contours are defined
with respect to the best-fit model. For instance, in
Ref.~\cite{Hannestad:2003ye}, the best-fit models for the {\it
degenerate} and {\it 1+r} cases correspond to high values of ($M$,
$N_{\rm eff}$) and are different from each other. This explains why
the authors find different contours in the two cases even in the
massless limit. The same occurs in our analysis when we do not
impose any prior. However, when at least the HST prior is taken into
account, the best-fit model is very close to $(M,
N_{\rm eff})=(0,3)$ and the contours only differ at high values of $M$ and
$N_{\rm eff}$.  As expected from the physical discussion in Sec.\
\ref{effects}, the model with only one massive neutrino is less
constrained: remember that in that case, for a fixed $M$ the
free-streaming wavenumber $k_{\rm FS}$ is larger. Therefore, the
damping of the matter power spectrum is less efficient.  The
one-dimensional bounds on the effective number of neutrinos
(marginalized over $M$) now reads $1.6 < N_{\rm eff} < 8.5$ at
$2\sigma$, and the limit on the total neutrino mass (marginalized over
$N_{\rm eff}$) increases to $M < 1.5$ eV (2$\sigma$).
The bounds for fixed integer values of $N_{\rm eff}$ are given in the
last column of Table \ref{table1}.

\section{Conclusions}
\label{conc}

We have calculated cosmological bounds on neutrino masses and
relativistic relics ($N_{\rm eff}$) using the latest data on CMB (WMAP
and ACBAR) and LSS (SDSS and 2dF galaxy surveys) in the framework of
an extended flat $\Lambda$CDM. In the cases in which a comparison is
possible, our results are in good agreement with those of previous
analyses \cite{Barger:2003vs}-\cite{Hannestad:2003ye}.
%\cite{Barger:2003vs,Tegmark:2003ud,Cuoco:2003cu,Allen:2003pt,
%  Barger:2003zg,Elgaroy:2003yh,Hannestad:2003xv,Pierpaoli:2003kw,
%  Crotty:2003th,Spergel:2003cb,Hannestad:2003ye}.  
In the well motivated case of three flavor neutrino with degenerate
masses, we found an upper limit on the total masses of $M < 1.0 ~({\rm
  resp.}~ 0.6)$ eV using only CMB and LSS data and priors (resp.
including priors on $h$ and $\Omega_{\Lambda}$).  The bound for four
thermalized neutrinos with only one of them carrying a significant
mass is $M < 0.8-1.2$ eV, depending on the priors used. Therefore, the
4-neutrino solution to the LSND results is not completely ruled out,
but some tension with cosmological data exists, especially if the
strong SN03 prior is taken into account.

In the case of arbitrary $N_{\rm eff}$,
our results are summarized in Fig.\ \ref{fig1abcd} and listed in Table
\ref{table1}. They clearly show the existence of a parameter degeneracy
between the total neutrino mass and $N_{\rm eff}$, a trend already
observed in previous works
\cite{Lesgourgues:2001he,Elgaroy:2003yh,Hannestad:2003xv,Hannestad:2003ye}
that we have explained in Sec.\ \ref{results}. 
External priors on $h$ and $\Omega_{\Lambda}$ are found to be of particular
importance for constraining respectively $N_{\rm eff}$ and $M$.

Since the standard $\Lambda$CDM model (with its three effectively
massless thermal relics) sits within the 1$\sigma$ preferred region,
we find no evidence for exotic physics such as out-of-equilibrium
neutrino decoupling, non-standard nucleosynthesis, extra relativistic
relics, a significant amount of hot dark matter, etc. However, large
deviations from the standard case are still compatible with
observations: for instance, a model with one neutrino of mass $M=1.5$
eV and eight relativistic degrees of freedom is allowed by CMB and LSS
data, even when all priors are included (see Fig.\ \ref{fig_split}).
In order to exclude this model, it is necessary to take into account
the prediction of standard BBN, which gives stronger limits on $N_{\rm
  eff}$.

The bounds obtained in this paper are based on the observation of
cosmological perturbations (CMB and LSS), combined with constraints on
the current expansion and acceleration rates of the Universe (HST and
SN priors). Therefore they are completely independent from the
predictions of primordial nucleosynthesis.  It is remarkable that in
the space of the two standard BBN free parameters $(\omega_b, N_{\rm
  eff})$, the preferred regions deduced from cosmological
perturbations and from primordial abundances are perfectly compatible
with each other, and more or less orthogonal: indeed, our analysis
(with the most restrictive priors) gives $0.0215 < \omega_b < 0.0235$
and $1.6 < N_{\rm eff} < 8.5$, while standard BBN favors $0.017 <
\omega_b < 0.026$ and $1.6 < N_{\rm eff} < 3.6$ \cite{Cuoco:2003cu}
(all these bounds are at the 2$\sigma$ level).

In order to test their robustness, we have also calculated the bounds
on $M$ and $N_{\rm eff}$ in the presence of primordial tensor
perturbations. Our results show that the bounds are practically
unchanged, because current cosmological data is able to distinguish
between the respective effects of tensors and neutrinos.
 
Finally, we have considered the impact of a different splitting of the
total neutrino mass among the individual states, an analysis also
recently performed in ref.\ \cite{Hannestad:2003ye}. We compared the
case of complete mass degeneracy (all neutrinos with the same mass)
with that where one neutrino state effectively possesses the whole
mass. We found that the bounds on the degenerate case are more
restrictive due to its more efficient free-streaming, in particular
for larger values if $N_{\rm eff}$. However, for three or four
neutrinos the differences between the two cases are not significant.
 
Our bounds are a clear indication that present cosmological data
provide interesting bounds on the neutrino sector, complementary to
those from terrestrial experiments. These include tritium beta decay
experiments, which provide a current upper bound on the total neutrino
mass of $6.6$ eV at 2$\sigma$ \cite{Bonn:tw}, while the KATRIN
experiment \cite{Osipowicz:2001sq} is planned to have an accuracy of
the order $0.35$ eV. Sub-eV sensitivity to neutrino masses is also
expected for experiments measuring neutrinoless double beta decays
\cite{Elliott:2002xe}, but only for Majorana neutrinos and with a
dependence on the details of the mixing matrix.

However, the cosmological bounds should be taken with care, due to
their dependence on the data (or priors) used, and also on the
assumption of a particular underlying model. Examples are given
  by the works \cite{Blanchard:2003du,Allen:2003pt}, where non-zero
  neutrino masses are preferred. This warning should not prevent us
to be confident on the power of future cosmological experiments to
limit (and eventually detect) neutrino masses and other neutrino
properties. For instance, forecast analyses have shown that with
future data there will be potential sensitivities to $\Delta N_{\rm
  eff}\sim 0.2$ \cite{Bowen:2001in,Bashinsky:2003tk} (eventually
improving BBN results) and neutrino masses of the order $0.1-0.2$ eV
with Planck and final SDSS data
\cite{Eisenstein:1998hr,Lesgourgues:1999ej,Hannestad:2002cn}, or with
galaxy and CMB lensing \cite{Kaplinghat:2003bh,Abazajian:2002ck}.

\section*{Acknowledgments}
This research was supported by a CICYT-IN2P3 agreement. SP was
supported by the Spanish grant BFM2002-00345, the ESF network Neutrino
Astrophysics and a Ram\'on y Cajal contract of MCyT.

\end{document}